\documentclass[aps,pr,twocolumn,superscriptaddress,groupedaddress]{revtex4}  
\usepackage{graphicx}  
\usepackage{dcolumn}   
\usepackage{bm}        
\usepackage{amssymb}   
\usepackage{amsmath}
\usepackage{ascmac}
\usepackage{color}
\begin{document}

\title{Measurement-based quantum computation utilizing the graph states of Bose-Einstein condensates and continuous variables}
\input 
\author{Genji Fujii \\Department of Nuclear Engineering, Kyoto University, 6158540 Kyoto, Japan\\}
\date{\today}

\begin{abstract}
Measurement-based quantum computation (MBQC) is a protocol for quantum computation that represents a model distinct from the circuit-based approach. MBQC has been proposed not only for qubits but also for qudits,  continuous-variable (CV) qubits, and Bose-Einstein condensates (BECs) qubits. In qubit-based MBQC, arbitrary rotations on the Bloch sphere can be performed by measuring a graph state. This naturally raises the question of whether arbitrary rotations on the Bloch sphere can similarly be achieved through measurements in other types of quantum bits. We have demonstrated that this can indeed be realized for BECs qubits by considering composite graph states involving CV qubits and BECs qubits.
\end{abstract}

\maketitle


\section{introduction}
 Quantum computers perform calculations in a Hilbert space spanned by $\{|0\rangle, |1\rangle\}$, known as a qubit. The quantum state of a qubit can be represented by its position on the Bloch sphere. The ability to perform arbitrary rotations on the Bloch sphere, i.e., to manipulate any quantum state of a qubit, ensures that quantum computers are powerful computational devices.
In addition to qubits, there is growing interest in continuous-variable quantum bits (CV qubits) based on photons. Photons are widely studied as a quantum communication medium with low decoherence over long distances \cite{1,2}, and it has been shown that universal quantum computation can be performed by CV qubits \cite{3,4}. Furthermore, photons have several advantages, such as enabling specific types of information processing that standard qubits cannot achieve, maintaining quantum coherence at room temperature, and being suitable for high-speed information processing. However, they also present challenges, such as the difficulty of implementing nonlinear transformations. Notably, there are also experimental studies on the creation of graph states using photons \cite{5,6,7,8,9}.\\
 \indent In quantum computation protocols, Bose-Einstein condensates (BECs) quantum computing has been proposed \cite{10}. Typically, when dealing with high-dimensional Hilbert spaces, such as those for qudits or qutrits, the Bloch sphere becomes more complex, making it visually challenging to ascertain whether arbitrary Bloch sphere rotations are feasible. In the case of BECs qubits, it looks like, despite the high-dimensional Hilbert spaces, quantum states can be represented on a single Bloch sphere. This makes the consideration of arbitrary rotations on the Bloch sphere particularly meaningful.
Also, BECs quantum computation has seen theoretical advancements in recent years \cite{11,12,13,14,15,16}, and particularly relevant to our study is the quantum repeater protocol \cite{11}. In Ref \cite{11}, $S^{z}S^{z}$ interactions are performed at a magic time to entangle atomic ensembles. The quantum repeater is then achieved by measuring intermediate nodes. However, in the quantum repeater protocol, quantum operations achieved through measurements are constrained to a discrete range. This is because the measurement results yield a Kronecker delta function.\\
\indent We referred to circuit-based quantum computation above. As another quantum computing protocol, measurement-based quantum computation (MBQC) has been proposed \cite{17,18}. MBQC is distinct from circuit-based quantum computation, as it performs quantum computation by measuring a graph state. Beginning with protocols using qubits \cite{17,18}, MBQC has been extended to those using qudits \cite{19,20,21}, CV qubits \cite{22,23}, and BECs qubits \cite{24}.
An intriguing question in MBQC is whether universal quantum computations can be implemented. For qubits, studies have investigated the types of graph structures that allow universal quantum computation, and, for instance, it is known that universal quantum computations cannot be performed on using a tree graph \cite{25}. Similarly, universal quantum computation has been shown to be possible with CV qubits \cite{23}. For MBQC using BECs qubits \cite{24}, universal quantum computation has been restricted to logical BECs qubits and has not been extended to arbitrary rotation on the BECs Bloch sphere.\\
\indent In this paper, we theoretically realized arbitrary rotation on the BECs Bloch sphere using composite graph states consisting of BECs qubits and CV qubits. However, to achieve universal quantum computation, not only arbitrary rotations on the BECs Bloch sphere but also two-qubit gates, such as the CZ gate, are required. This will be discussed later. The fundamental idea is the same as in MBQC: prepare a graph state and perform quantum computation through measurements. What distinguishes this approach from other MBQC protocols is the consideration of graph states that mix two types of qubits, namely BECs qubits and CV qubits.
As a result, continuous variables emerge, and it is noteworthy that Dirac delta functions arising due to measurements play a critical role. In the quantum repeater protocol \cite{11} mentioned above, when measuring a intermediate node, Kronecker delta functions appear, leading to discrete quantum operations. In contrast, our framework enables continuous quantum operations by integrating over the CV qubits to resolve Dirac delta divergences. We will also discuss the feasibility of arbitrary rotation on the Bloch sphere for representative graph structures. 
 \section{BECs qubits and CV qubits}
  In this section, we review the notations used in previous studies for the systems we deal with. We introduce BECs and CV systems in quantum computation. In this paper, we call those systems BECs qubits and CV qubits. BECs qubits in the bosonic notation is
\begin{equation}
|\alpha,\beta\rangle\rangle\equiv\frac{1}{\sqrt{N!}}(\alpha a^{\dagger}+\beta b^{\dagger})^{N}|\rm{vac}\rangle,
\end{equation}
where $a^{\dagger}, b^{\dagger}$ obey commutation relations $[a,a^{\dagger}]=[b,b^{\dagger}]=1$.  $\alpha$ and $\beta$ are satisfying $|\alpha|^{2}+|\beta|^{2}=1$ and $|\rm{vac}\rangle$ is a vacuum state. Eq. (1) is also called spin coherent state \cite{10} or spinor state \cite{15}. Further, the spin operators are defined as
\begin{equation}
\begin{split}
S^{x}&=a^{\dagger}b+b^{\dagger}a,\\
S^{y}&=-ia^{\dagger}b+ib^{\dagger}a,\\
S^{z}&=a^{\dagger}a-b^{\dagger}b,\\
N&=a^{\dagger}a+b^{\dagger}b\\
&=n^{a}+n^{b}.
\end{split}
\end{equation}
Spin operators satisfy the SU(2) commutation relation $[S^{i},S^{j}]=2i\epsilon_{ijk}S^{k}$, where $\epsilon_{ijk}$ is the Levi-Civita antisymmetric tensor $i, j, k = x, y, z$. The Fock basis, which is the eigenstate of $S^{z}|k\rangle=(2k-N)|k\rangle$, is defined as
\begin{equation}
\begin{split}
|k\rangle \equiv\frac{(a^{\dagger})^{k}(b^{\dagger})^{N-k}}{\sqrt{k!(N-k)!}}|vac\rangle.
\end{split}
\end{equation}
On the other hand, CV systems or a CV aubit is defined as
\begin{equation}
|\psi\rangle=\int_{-\infty}^{\infty}\psi(x)|x\rangle dx,
\end{equation}
where $x$ is a position and $\psi(x)$ is a probability amplitude related to $x$. CV qubits can also be represented by momentum notations
\begin{equation}
|\psi\rangle=\int_{-\infty}^{\infty}\psi(p)|p\rangle dp,
\end{equation}
where $p$ is a momentum and $\psi(p)$ is a probability amplitude related to $p$. $\hat{x}$ and $\hat{p}$ are position and momentum operators, which satisfies the following relation $[\hat{x},\hat{p}]=i$. Here we define $\hat{x}=\frac{\hat{A}+\hat{A}^{\dagger}}{2}$ and $\hat{p}=\frac{\hat{A}-\hat{A}^{\dagger}}{2i}$, $\hat{A}$ and $\hat{A}^{\dagger}$ are Harmonic oscillator creation and  annihilation operators. The eigenstates of $\hat{x}$ and $\hat{p}$ are
\begin{equation}
\begin{split}
\hat{x}|x\rangle&=x|x\rangle\\
\hat{p}|p\rangle&=p|p\rangle.
\end{split}
\end{equation}
 \section{Formalism}
 In this section, we define the graph state in a composite system of BECs qubits and CV qubits, and discusses rotation on the BECs Bloch sphere by measurements.
 \subsection{Definition of BECs and CV graph states}
\includegraphics[width=80mm]{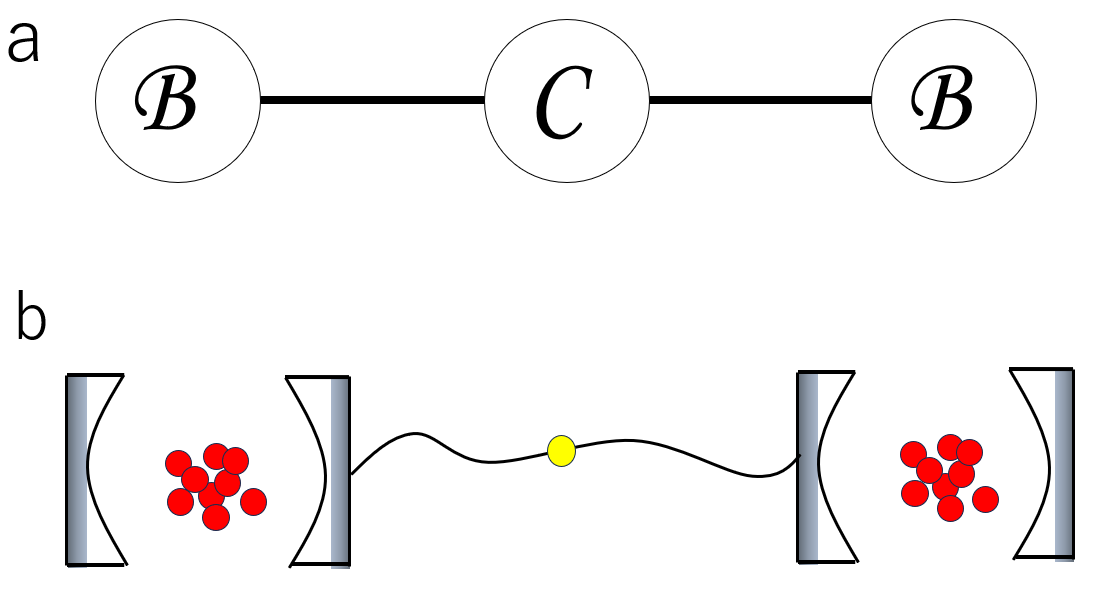}\\
\\
\ \  FIG. 1: (a) The graph state with BECs qubits and CV qubits. The vertex labeled $B$ represents the BECs qubit, and $C$ represents the CV qubit. The edges are represented by a CZ gate. (b) Possible experimentally implementation involves atomic ensembles in cavities connected via optical fibers. BECs qubits, polarized in the $S^{x}$ direction, are entangled with photons, which serve as CV qubits.\\
\\
\includegraphics[width=80mm]{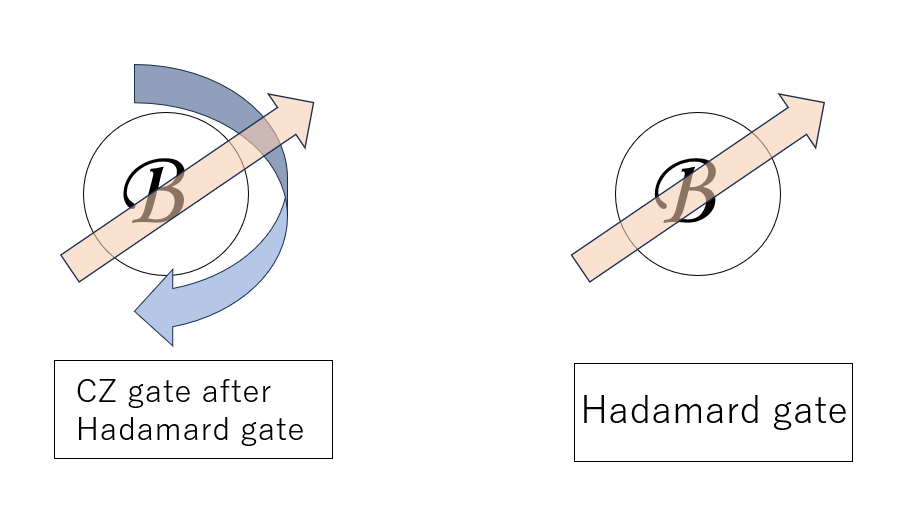}\\
\\
\ \  FIG. 2: A useful representation in the context of our structure. The left figure shows application of Hadamard gate before the application of the CZ gate, omited edges and other vertices in this figure. the right figure shows the application of Hadamard gate for the vertex of single BECs qubit.\\
\\
 We now define a graph state that combines BECs qubits and CV qubits. Two-BECs qubits and one-CV qubit graph state is shown in Fig. 1. The vertices with CV qubits labeled as $C$ and BECs qubits labeled as $B$. The edges are represented by a CZ gate that connect the CV qubits and BECs qubits. In our construction, the CZ gate Hamiltonian as
 \begin{equation}
 H=n_{1}^{b}\hat{x_{2}}=\frac{1}{4}(N_{1}-S^{z}_{1})(A^{\dagger}_{2}+A_{2}).
 \end{equation}
 This Hamiltonian time evolution for a time $t$ between BECs qubits and CV qubits gives the unitary evolution
 \begin{equation}
 \begin{split}
& e^{-iHt}|\frac{1}{\sqrt{2}}, \frac{1}{\sqrt{2}}\rangle\rangle_{1}\int_{-\infty}^{\infty}\psi(x)|x\rangle_{2}dx\\
 &=\int_{-\infty}^{\infty}|\frac{1}{\sqrt{2}}, \frac{e^{-ixt}}{\sqrt{2}}\rangle\rangle_{1}\psi(x)|x\rangle_{2}dx.
 \end{split}
 \end{equation}
 Unlike the CZ gate discussed in Ref. \cite{24}, this type of CZ gate produces an entangled state between the atomic ensemble and the continuous variable. Notably, so continuous variables are represented as a superposition of arbitrary real numbers $x$, the phase component of the BECs qubit is a continuous value. We note that the choice of $x$ or $p$ is free unless special attention is required, such as in the case of the GKP code \cite{26} if one considers error correction codes.\\
\indent We also introduce a representation that is useful for visually expressing the graph state quantum computating. Fig. 2 in left illustrates the application of Hadamard gate before the application of the CZ gate. Note that we omited edges and other vertices in this figure. right figure  illustrates Hadamard gate acts on the BECs qubit. These are the notations used in this paper for theoretical construction.
 \subsection{Measurement graph states and arbitrary rotation on the BECs Bloch sphere}
 In this section, we investigate the measurement of graph states and then arbitrary rotations on the BECs Bloch sphere. In the previous section, we constructed the graph states using BECs qubits and CV qubits, which are possible to perform rotations along the $z$-axis and $x$-axis by consuming two resources (Fig. 3). Furthermore, we will show that arbitrary rotations on the BECs Bloch sphere can be implemented by consuming four resources in three-BECs and two-CV qubits graph state (Fig. 4). First, let us consider rotations about the $z$-axis. The outline of the scheme is shown in Fig. 3. The two-BECs, one-CV graph state, in the case of (a), is
\begin{equation}
\begin{split}
|G\rangle\rangle_{z}&=\int_{-\infty}^{\infty}|\frac{1}{\sqrt{2}}, \frac{e^{-xi}}{\sqrt{2}}\rangle\rangle_{1}\psi(x)|x\rangle_{2}\\
&\otimes|\frac{1}{\sqrt{2}}, \frac{e^{-2\pi xi/L}}{\sqrt{2}}\rangle\rangle_{3}dx,
\end{split}
\end{equation}
where we choice the ``magic time'' time evolution $t=2\pi /L$ \cite{11} between the CV qubit 2 and the BECs qubit 3, where we select the $L$ is integer and the large number and time evolution $t=1$ between BECs qubit 1 and CVqubit 2. On the other hand, the graph state, in the case of Fig. 3 (b), is
\includegraphics[width=70mm]{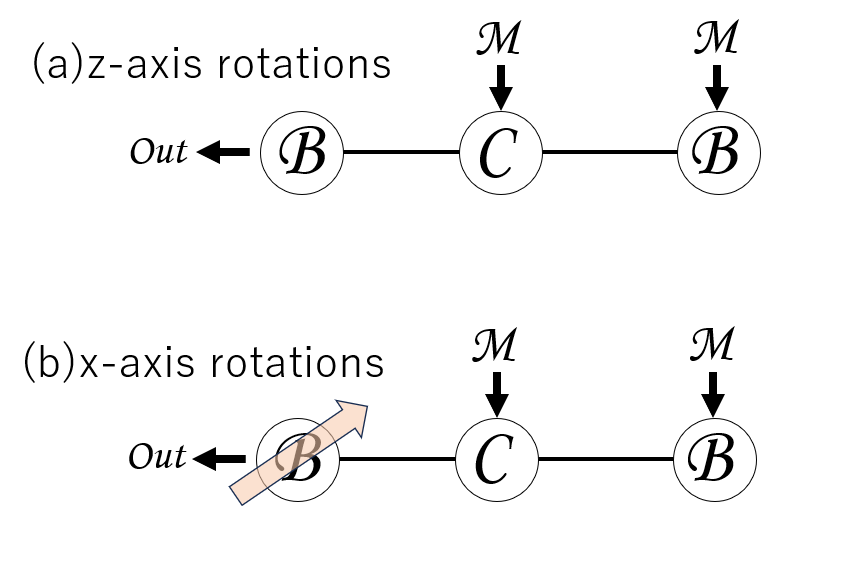}\\
\\
\ \  FIG. 3: Measurement-induced rotations on the BECs Bloch sphere: (a) $z$-axis rotation, (b) $x$-axis rotation. $M$ represents the measurement, and the leftmost BECs qubit is the outcome state.\\
\\
\includegraphics[width=80mm]{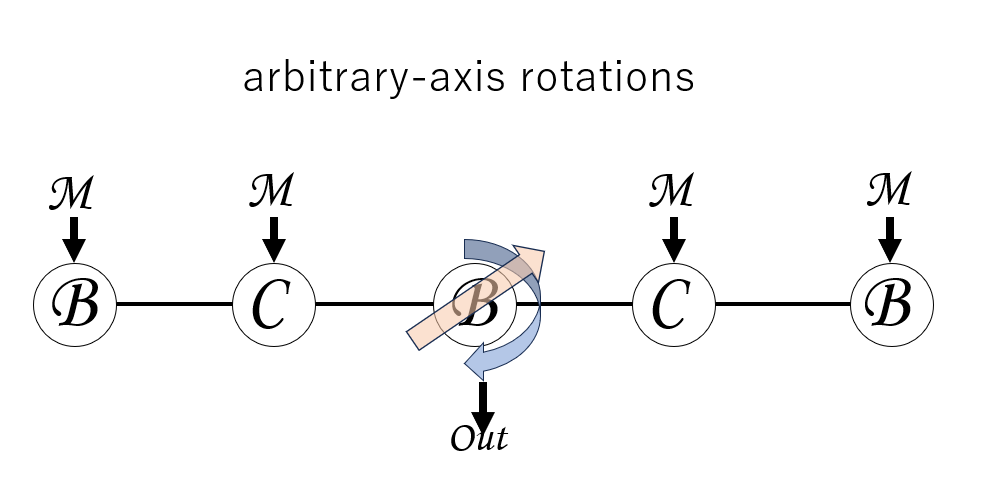}\\
\\
\ \  FIG. 4: Measurement-induced arbitrary rotation on the BECs Bloch sphere. By consuming four resources, then the central BECs qubit is the arbitrary rotation outcome state.\\
\begin{equation}
\begin{split}
|G\rangle\rangle_{x}&=\int_{-\infty}^{\infty}|\cos(x), i\sin(x)\rangle\rangle_{1}\psi(x)\\
&\otimes|x\rangle_{2}|\frac{1}{\sqrt{2}}, \frac{e^{-2\pi ix/L}}{\sqrt{2}}\rangle\rangle_{3}dx.
\end{split}
\end{equation}
Since the measurement effects on these two types of graph states are the same, we mainly discuss eq. (9) case. In order to achieve any $z$-axis rotation, first, we measure the BECs qubit 3 (The rightmost) in the wanted phase $x$-basis. This $x$-axis measurement operator is expressed as
\begin{equation}
M_{\theta}^{(x)}=|\theta^{(x)}_{3}\rangle\langle\theta^{(x)}_{3}|,
\end{equation}
where
\begin{equation}
\begin{split}
|\theta_{3}^{(x)}\rangle&=e^{-in^{b}_{3,l}(2\pi\theta/L-\pi/2)}e^{-in^{b}_{3,m}(2\pi\theta/L-\pi/2)}\\
&\times e^{-3iS^{y}\pi/4}|k\rangle,
\end{split}
\end{equation}
 which is an unnormalized state. We note that the state in which Hadamard gate $(H= e^{-3iS^{y}\pi/4})$ acts on the Fock basis is $(a^{\dagger}+b^{\dagger})^{k}(a^{\dagger}-b^{\dagger})^{N-k}/\sqrt{k!(N-k)!}|{\rm vac}\rangle$. $n^{b}_{3,l}$ and $n^{b}_{3,m}$ act as $n^{b}_{3,l}|l\rangle=(k-l)|l\rangle$, $n^{b}_{3,m}|m\rangle=(N-k-m)|m\rangle$, where  $|l\rangle=(a^{\dagger})^{l}(b^{\dagger})^{k-l}/\sqrt{l!(k-l)!}|{\rm vac}\rangle$ and $|m\rangle=(a^{\dagger})^{m}(b^{\dagger})^{N-k-m}/\sqrt{m!(N-k-m)!}|{\rm vac}\rangle$. The results of the measurements for BECs 3 at this basis are
 \begin{equation}
 \begin{split}
&\langle\theta^{(x)}_{3} |\frac{1}{\sqrt{2}}, \frac{e^{-2\pi ix/L}}{\sqrt{2}}\rangle\rangle_{3}\\
&=\sqrt{\frac{N!}{q_{3}!(N-q_{3})!}}\cos^{q_{3}}(\pi(x-\theta)/L+\pi/4)\\
&\times\sin^{N-q_{3}}(\pi(x-\theta)/L+\pi/4)\\
&=\sqrt{\frac{N!}{q_{3}!(N-q_{3})!}}\cos^{N}(\pi(x-\theta)/L+\pi/4)\\
&\times\tan^{N-q_{3}}(\pi(x-\theta)/L+\pi/4),
\end{split}
 \end{equation}
 where $q_{3}$ is a measurement result. Also we neglect grobal phases. Furthermore, we consider the cosine addition formula
 \begin{equation}
 \begin{split}
 &\cos^{N}(\pi(x-\theta)/L+\pi/4)\\
 &=\frac{1}{\sqrt{2^{N}}}(\cos(\pi(x-\theta)/L)-\sin(\pi(x-\theta)/L))^{N},
 \end{split}
 \end{equation}
 which can be approximated by an exponential function.
 \begin{equation}
 \begin{split}
( \cos(\theta)-\sin(\theta))^{N}&\approx \exp((-\theta-\theta^{2}/2)N),\\
&=\exp(-\theta N)\exp(-\theta^{2}N/2)
\end{split}
 \end{equation}
 which is valid for $\theta\ll1$. In the eq. (14) case, $\pi(x-\theta)\ll L$ is required. Therefore, the squared term of the exponential function in eq. (14) can be replaced with the Dirac delta function.
 \begin{equation}
 \exp(\pi^{2}(x-\theta)^{2}N/2L^{2})=\sqrt{\frac{\pi}{N}}\delta(\frac{\pi}{L}(x-\theta)),
 \end{equation}
 where we defined the Dirac delta function bellow notation. the Dirac delta function as
 \begin{equation}
 \delta(x-y)=\lim_{\epsilon\rightarrow 0}\frac{1}{\sqrt{\pi \epsilon}}\exp(-(x-y)^{2}/\epsilon).
 \end{equation}
 Eq. (16) is valid for $\pi^{2}/2L^{2}\ll N$. so, $N\rightarrow \infty$ is required. Therefore, the state after measuring $|G\rangle\rangle_{z}$ is
 \begin{equation}
 \begin{split}
 |G\rangle\rangle_{z}'&=\sqrt{\frac{N!}{q_{3}!(N-q_{3})!}}\sqrt{\frac{L}{2^{N}N}}\int_{-\infty}^{\infty}|\frac{1}{\sqrt{2}}, \frac{e^{-ix}}{\sqrt{2}}\rangle\rangle_{1}\\&\times\psi(x)|x\rangle_{2}\delta(x-\theta)\tan^{N-q_{3}}(\pi(x-\theta)/L+\pi/4)\\
 &\times\exp(\pi(x-\theta)/L)dx.
 \end{split}
 \end{equation}
 Now, executing the integration,
  \begin{equation}
 |G\rangle\rangle_{z}'=\sqrt{\frac{N!}{q_{3}!(N-q_{3})!}}\sqrt{\frac{L}{2^{N}N}}|\frac{1}{\sqrt{2}}, \frac{e^{-i\theta}}{\sqrt{2}}\rangle\rangle_{1}\psi(\theta)|\theta\rangle_{2}.
 \end{equation}
 Finally, we measure the CV qubit 2. This measurement process for CV qubits can be performed by homodyne detection. The BECs qubit 1 is output as a state rotated by an arbitrary angle $\theta$ on the $x$-$y$ plane of the BECs Bloch sphere through measurement. By considering a similar measurement for graph state $|G\rangle\rangle_{x}$, arbitrary rotations around the $x$-axis can be realized. Measurements for arbitrary rotations on the BECs Bloch sphere (Fig. 4) are performed in a similar manner, however, we require careful attention to the order in which Hadamard gate and the CZ gate are applied during the preparation of the graph state. a procedure for preparing the graph state is explained bellow.\\
 \\
 1.~Apply two consecutive CZ gates to the left three vertices.\\
 \\
 2.~Apply to Hadamard gate on the central BECs qubit.\\
 \\
 3.~Apply two consecutive CZ gates to the right three vertices.\\
 \\
 \indent With the above steps, the preparation of the graph state capable of performing arbitrary rotations on the BECs Bloch sphere is feasible.\\
 \indent We also note the following important point: For the measurement of the state applying on Hadamard gate on BECs qubit on the right side of Fig. 2, performing the measurement in the following basis enables quantum computation
 \begin{equation}
 M_{\theta}^{(H)}=|\theta^{(H)}\rangle\langle\theta^{(H)}|,
 \end{equation}
 where $|\theta^{(H)}\rangle=e^{-3iS^{y}\pi/4}|\theta^{(x)}\rangle$. It should also be noted that the state shown BECs qubit, in the left side of Fig. 2, is exclusively for output.
 \section{Quantum computation on several graphs}
 \includegraphics[width=80mm]{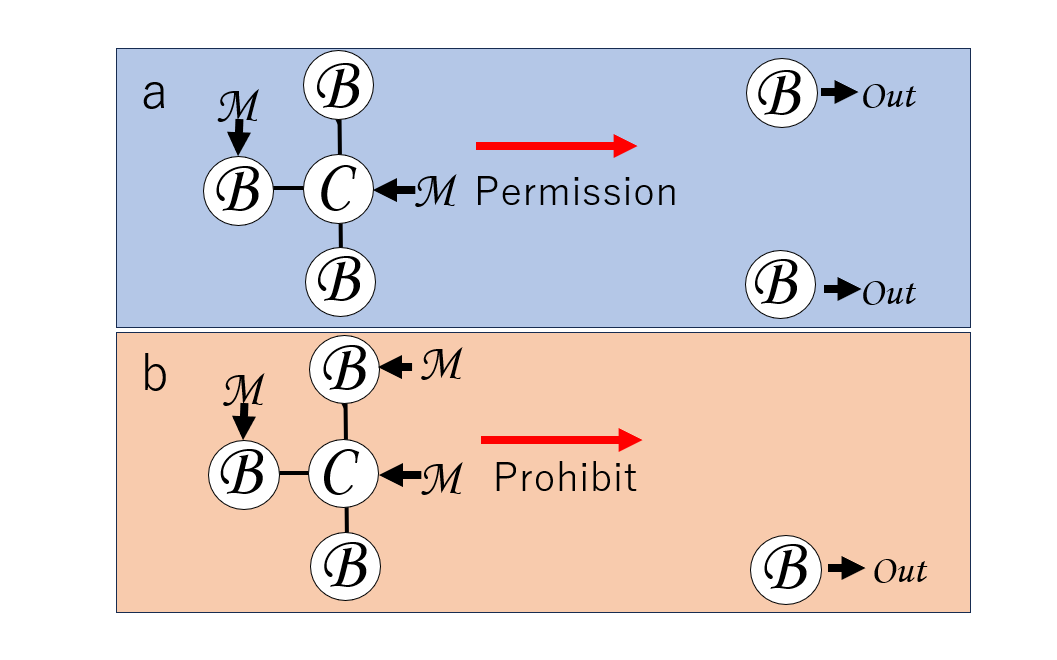}
\ \  FIG. 5: (a). For each measured BECs qubits, there must be exactly one CV qubit connected by an edge.
(b). For each CV qubit connected by an edge, there are two BEC qubits being measured, making this operation prohibited. This is because two Dirac delta functions would appear, which cannot be resolved with a single integration of the CV qubit.\\
\\
\\
 \includegraphics[width=80mm]{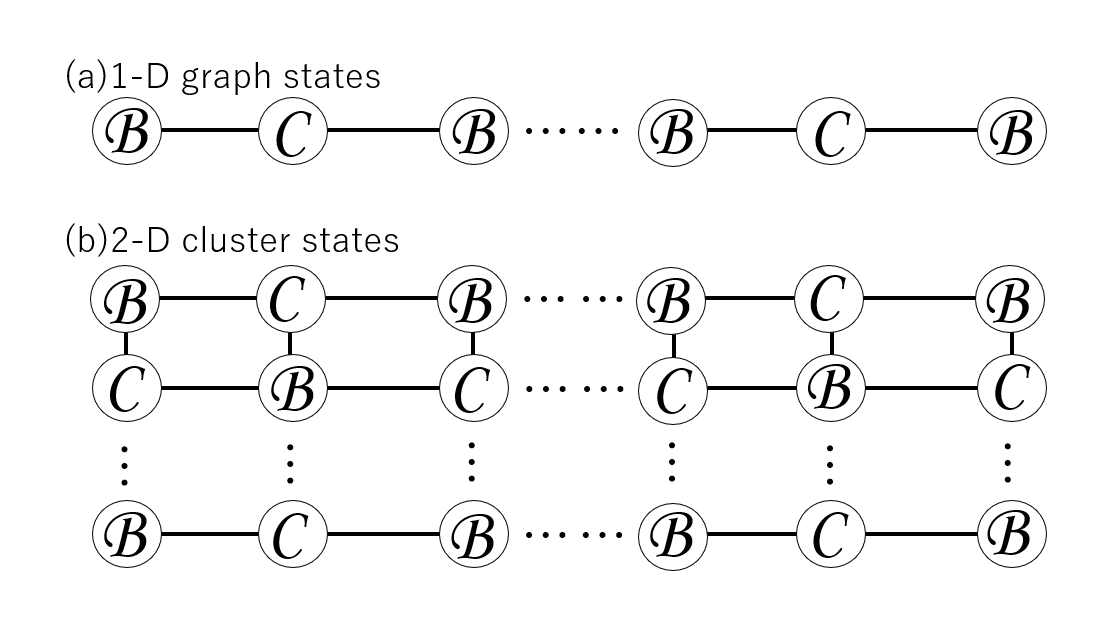}
 \includegraphics[width=80mm]{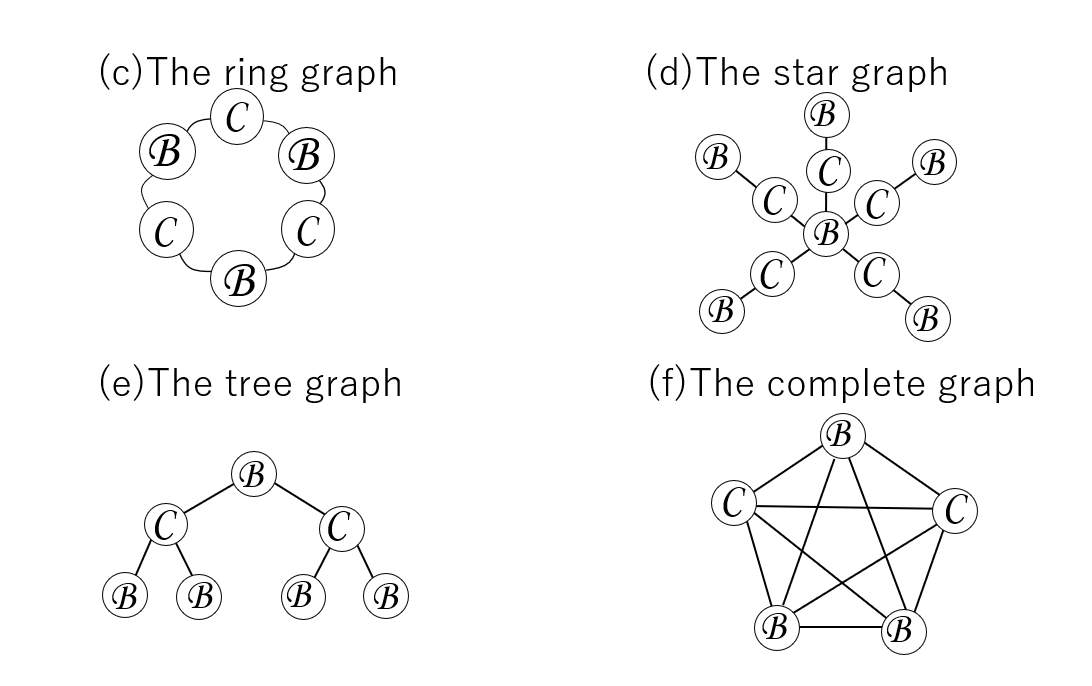}\\
\\
\ \  FIG. 6: (a)$\sim$(f): Sketches of representative graph structures\\
\\
\\
 \indent In our framework, it is an intriguing question to determine what types of graph connections are allowed and what shapes of graphs enable arbitrary rotation on the BECs Bloch sphere. We consider six representative graphs. Although not explicitly shown in Fig. 6, following discussion, we assume that, if necessary, Hadamard gate applied appropriately for $x$-axis rotations and arbitrary rotations. First, our graph state has several constraints:\\
\\
1.~The neighboring vertex connected to vertex B must be C. In other words, a graph where C is adjacent to C or B is adjacent to B does not exist.\\
\\
2.~The number of BECs qubits to be measured must equal the number of connected CV qubits. This is because the Dirac delta functions contributed by the measurement of the BECs qubits must be canceled out by the integration over the CV qubits (Fig. 5).\\
\\
3. Considering the total number of output and measured BECs qubits, the number of BECs qubit vertices, $N_{B}$, and the number of CV qubit vertices, $N_{C}$, in the graph state must satisfy the following condition: $N_{C}\leq N_{B}$\\
\\
\\
(a). 1-D graph states\\
\\
In this type of graph, there are two possible cases: one where the BECs qubits at the ends serve as outputs, and another where the BEC qubits that are not at the ends serve as outputs. It is evident that the former can only perform rotations around either the $z$-axis or the $x$-axis, whereas the latter has the potential to perform arbitrary rotations.\\
\\
(b). 2-D cluster states\\
\\
In this type of graph, performing a measurement on any of the BECs qubits violates Rule 2 mentioned above, making quantum computation impossible as is. However, it may become possible to perform quantum computation by measuring some of the CV qubits to create gaps beforehand.
\\
\\
(c). The ring graph\\
\\
The ring graph is a graph in which every vertex has exactly two adjacent vertices, and consecutive edges form a closed loop. In this graph,  measuring the BECs qubit at any vertex of this graph is prohibited as it violates Rule 2. However, by measuring the CV qubit at a certain vertex to open the loop, the graph transforms into a 1-D graph state, enabling quantum computation.\\
\\
(d). The star graph\\
\\
The star graph is a type of graph consisting of a central node and multiple nodes connected to it. In this graph, by considering a measurement flow from the peripheral nodes to the central node, arbitrary rotations can be performed on the BECs Bloch sphere of the central BECs qubit. On the other hand, starting measurements from the central BECs qubit is prohibited as it violates Rule 2.\\
\\
(e). The tree graph\\
\\
The tree graph is an undirected graph that is connected and contains no cycles. In this graph, measurements cause an upward flow. When one of the leaf BECs qubits is measured, the measurement result flows to the CV qubit node connected to that BECs qubit node. As a result, rotations around the $z$-axis or $x$-axis occur for the other leaf BECs qubits, and arbitrary rotations can be induced on the root BEC qubit. On the other hand, a downward flow starting from the root BECs qubit is prohibited by Rule 2.\\
\\
(f). The complete graph\\
\\
The complete graph is an undirected graph in which all vertices are connected to each other. In this graph, if the number of nodes is 3 or more, it violates Rule 1 and is therefore prohibited existing.
 \section{Conclusion and Discussion}
  We proposed a protocol for arbitrary rotations on the BECs Bloch sphere by considering graph states in the composite system of BEC qubits and CV qubits. We also investigated whether arbitrary rotations are possible for several representative graph structures. However, one of the criteria for quantum computers to demonstrate powerful computational capabilities is the ability to perform universal quantum computation. This requires not only arbitrary rotations on the Bloch sphere but also two-qubit gates such as the CZ gate.\\
  \indent The concept of universal quantum computation with BECs qubits is intricate. First, in the case of BECs qubits, in addition to the BECs Bloch sphere representation, there is also a representation based on the Fock basis, where they can be regarded as a type of qudits. However, regarding the BECs Bloch sphere representation, it is an SU(2) group representation, suggesting that universal quantum computation could be achievable with CZ gates in graph states, as shown in previous studies \cite{24}, or by possibly implementing CZ gates using repeater protocols \cite{11}. It should also be noted that when introducing these concepts, Rule 1 from Sec. IV must be relaxed, and entanglement between BECs qubits must also be taken into consideration. In that case, the shape of the graph that can perform universal quantum computation is left as an open problem.\\
  \indent We also comment on decoherence. The main decoherence factors in CV qubit and BECs are particle loss and dephasing. Several error correction codes for these have been studied \cite{26,27,28,29}. For example, the GKP code \cite{26} in CV qubit is currently a promising error correction code for particle loss and dephasing. In BECs systems, particle loss and dephasing are known to not scale to the number of particles \cite{10}, and specific error correction codes such as cat code \cite{27} may be applicable. The best combination of error correction codes in our system and the development of new error correction codes is also an open problem.
\input 


\begin{thebibliography}{99}
\bibitem{1}Pirandola, Stefano, et al. "Fundamental limits of repeaterless quantum communications." Nature communications 8.1 (2017): 1-15.
\bibitem{2}Mao, Yingqiu, Pei Zeng, and Teng‐Yun Chen. "Recent advances on quantum key distribution overcoming the linear secret key capacity bound." Advanced Quantum Technologies 4.1 (2021): 2000084.
\bibitem{3}Lloyd, Seth, and Samuel L. Braunstein. "Quantum computation over continuous variables." Physical Review Letters 82.8 (1999): 1784.
\bibitem{4}Takeda, Shuntaro, and Akira Furusawa. "Universal quantum computing with measurement-induced continuous-variable gate sequence in a loop-based architecture." Physical review letters 119.12 (2017): 120504.
\bibitem{5}Azuma, Koji, Kiyoshi Tamaki, and Hoi-Kwong Lo. "All-photonic quantum repeaters." Nature communications 6.1 (2015): 1-7.
\bibitem{6}Azuma, Koji, et al. "Quantum repeaters: From quantum networks to the quantum internet." Reviews of Modern Physics 95.4 (2023): 045006.
\bibitem{7}Menicucci, Nicolas C. "Temporal-mode continuous-variable cluster states using linear optics." Physical Review A—Atomic, Molecular, and Optical Physics 83.6 (2011): 062314.
\bibitem{8}Fukui, Kosuke, Warit Asavanant, and Akira Furusawa. "Temporal-mode continuous-variable three-dimensional cluster state for topologically protected measurement-based quantum computation." Physical Review A 102.3 (2020): 032614.
\bibitem{9}Azim, Ali Waqar, et al. "Dual-mode time domain multiplexed chirp spread spectrum." IEEE Transactions on Vehicular Technology (2023).
\bibitem{10}Byrnes, Tim, Kai Wen, and Yoshihisa Yamamoto. "Macroscopic quantum computation using Bose-Einstein condensates." Physical Review A—Atomic, Molecular, and Optical Physics 85.4 (2012): 040306.
\bibitem{11}Pyrkov, Alexey N., Ilia D. Lazarev, and Tim Byrnes. "Quantum repeater protocol for deterministic distribution of macroscopic entanglement." arXiv preprint arXiv:2408.00141 (2024).
\bibitem{12}Chaudhary, Manish, et al. "Remote state preparation of two-component Bose-Einstein condensates." Physical Review A 103.6 (2021): 062417.
\bibitem{13}Chaudhary, Manish, et al. "Macroscopic maximally-entangled-state preparation between two atomic ensembles." Physical Review A 108.3 (2023): 032420.
\bibitem{14}Chaudhary, Manish, et al. "Macroscopic quantum teleportation with ensembles of qubits." arXiv preprint arXiv:2411.02968 (2024).
\bibitem{15}Byrnes, Tim. "Multipartite spin coherent states and spinor states." Physical Review A 109.2 (2024): 022438.
\bibitem{16}Kitzinger, Jonas, et al. "Two-axis two-spin squeezed states." Physical Review Research 2.3 (2020): 033504.
\bibitem{17}Raussendorf, Robert, and Hans J. Briegel. "A one-way quantum computer." Physical review letters 86.22 (2001): 5188.
\bibitem{18}Raussendorf, Robert, Daniel E. Browne, and Hans J. Briegel. "Measurement-based quantum computation on cluster states." Physical review A 68.2 (2003): 022312.
\bibitem{19}Keet, Adrian, et al. "Quantum secret sharing with qudit graph states." Physical Review A—Atomic, Molecular, and Optical Physics 82.6 (2010): 062315.
\bibitem{20}Looi, Shiang Yong, et al. "Quantum-error-correcting codes using qudit graph states." Physical Review A—Atomic, Molecular, and Optical Physics 78.4 (2008): 042303.
\bibitem{21}Tang, Weidong, Sixia Yu, and C. H. Oh. "Greenberger-Horne-Zeilinger paradoxes from qudit graph states." Physical review letters 110.10 (2013): 100403.
\bibitem{22}Gu, Mile, et al. "Quantum computing with continuous-variable clusters." Physical Review A—Atomic, Molecular, and Optical Physics 79.6 (2009): 062318.
\bibitem{23}Menicucci, Nicolas C., et al. "Universal quantum computation with continuous-variable cluster states." Physical review letters 97.11 (2006): 110501.
\bibitem{24}Fujii, Genji. "Measurement-based quantum computation using two-component BECs." Physica Scripta 98.6 (2023): 065109.
\bibitem{25}Yoran, Nadav, and Anthony J. Short. "Classical simulation of limited-width cluster-state quantum computation." Physical review letters 96.17 (2006): 170503.
\bibitem{26}Gottesman, Daniel, Alexei Kitaev, and John Preskill. "Encoding a qubit in an oscillator." Physical Review A 64.1 (2001): 012310.
\bibitem{27}Leghtas, Zaki, et al. "Hardware-efficient autonomous quantum memory protection." Physical Review Letters 111.12 (2013): 120501.
\bibitem{28}Michael, Marios H., et al. "New class of quantum error-correcting codes for a bosonic mode." Physical Review X 6.3 (2016): 031006.
\bibitem{29}Cai, Weizhou, et al. "Bosonic quantum error correction codes in superconducting quantum circuits." Fundamental Research 1.1 (2021): 50-67.
\end{thebibliography}
\end{document}